# High-throughput nanoindentation mapping of cast IN718 nickel-based superalloy: influence of the Nb concentration

Alberto Orozco-Caballero[a, b], Cristina Gutierrez[a,c], Bin Gan[d] and Jon M. Molina-Aldareguia[a,*]

([a]) *IMDEA Materials Institute, C/ Eric Kandel 2, 28906, Getafe, Madrid, Spain*
([b]) *Now at: Department of Mechanical Engineering, Chemistry and Industrial Design, Escuela Técnica Superior de Ingeniería y Diseño Industrial - Polytechnic University of Madrid, Ronda de Valencia, 3, 28012, Madrid, Spain*
([c]) *Department of Materials Science and Engineering, Carlos III University, 28912 Leganes, Madrid*
([d]) *Beijing Key Laboratory of Advanced High Temperature Materials, Central Iron and Steel Research Institute, Beijing, 100081, China*

([*]) *Corresponding author: jon.molina@imdea.org.*

*Abstract*

A high-throughput correlative study of the local mechanical properties, chemical composition and crystallographic orientation has been carried out in selected areas of cast Inconel 718 specimens subjected to three different tempers. The specimens showed a strong Nb segregation at the scale of the dendrite arms, with local Nb contents that varied between 2 wt.% in the core of the dendrite arms to 8 wt.% in the interdendritic regions and 25 wt.% within the second phase particles (MC carbides, Laves phases and δ phase needles). The nanohardness was found to correlate strongly with the local Nb content and the temper condition. On the contrary, the indentation elastic moduli was not influenced by the local chemical composition or temper condition, but directly correlated with the crystallographic grain orientation, due to the high elastic anisotropy of nickel alloys.

# 1. Introduction

Cast and forged In718 polycrystalline Ni-base superalloys present high temperature strength and fatigue resistance even in oxidizing and corrosive environments [1]. Such performance makes this material suitable for use as turbine discs and some static components of aero-engines working at temperatures up to 600-700 ºC [2]. The formulation of this alloy includes various transition metals which lead to strengthening by solid solution, precipitation via several second phase particles, γ', γ'', δ ($Ni_3Nb$), MC carbides and Laves phases, as well as grain size refinement and a high twin boundaries density [3]. In casting processes, chemical segregation occurs during the solidification stage leading to local variations in the mechanical properties of the material at the microstructural scale that have not been studied before.

The recent advances in high-speed nanoindentation mapping methods, such as XPM (accelerated property mapping) [4], enable evaluating local variations in the mechanical properties of large areas with lateral resolutions of a few micrometres. Such resolution values are similar to those typically selected in electron microscopy techniques for determining the local composition and the crystallographic orientation, such as EDS (electron dispersive X-ray spectroscopy) and EBSD (electron back scatter diffraction), respectively. Hardness is usually determined as the ratio between the maximum indentation load and the projected area of the residual imprint [5]. Nevertheless, when the indentation depth ranges from nanometres to a few micrometres such task becomes tedious, especially in the case of high-speed nanoindentation maps where the aim is to measure thousands of indentations. This issue can be addressed by depth-sensing instrumented indentation, where the indentation load ($P$) and the penetration depth ($d$) of the indenter are continuously recorded. Provided that the geometry of the indenter is known, the hardness and elastic modulus can be directly inferred from the indentation load-penetration curves using the Oliver and Pharr method [6]. In the case of high-speed nanoindentation maps is, however, important to keep track of the indenter geometry during the process, because this can be altered by tip wear over the thousands of indentations involved in the maps.

In this study, we correlate the XPM maps with the crystallographic orientation obtained by EBSD and the compositional maps obtained by EDS of selected areas of cast Inconel 718 specimens subjected to three different tempers. The EDS results showed acute Nb segregation across the dendritic arms and in the interdendritic regions that lead to strong nanohardness gradients in the XPM maps. Nevertheless, the XPM maps show relatively constant elastic moduli inside the grains, not influenced by the local chemical composition but a direct dependence of the grain orientation, as expected for materials with high elastic anisotropy like nickel. Determining the dependence of the local mechanical properties with chemical segregation is key for tailoring the properties of the resulting material during casting or other solidification processes, such us welding, repairing or 3D printing, for optimizing processing parameters and for developing microstructure based models.

## 2.- Results

## 2.1.- Microstructure

The material used in the present study was a cast IN718 polycrystalline Ni-base superalloy, with the average composition shown in Table 1.

Table 1. Chemical composition of IN718 alloy (in wt.%)

| Cr | Fe   | Nb  | Mo | Ti  | Al  | Ni   |
|----|------|-----|----|-----|-----|------|
| 19 | 18.5 | 5.1 | 3  | 0.9 | 0.5 | Bal. |

The alloy was heat treated to three different temper conditions:

i. Solubilisation (S): 965 ºC for 1 hour followed by oil quenching.

ii. Peak aging precipitation (P): following solubilisation treatment, 720 ºC during 8 h followed by 2 h cooling time towards 620 ºC and then, 620 ºC during 8 h and air cooling.

iii. Overaging (O): after peak aging precipitation, 800 ºC during 36 h followed by furnace cooling.

Optical micrographs of the cast Inconel 718 alloy are presented in Fig. 1a and b. The microstructure was characterized by large irregular grains in the millimetre size range. As typically occurs during solidification, each grain develops by dendritic growth, leaving behind interdendritic regions decorated by coarse second phase particles, mainly δ ($Ni_3Nb$), MC carbides and Laves phases, as shown in Fig 1c. The volume fractions of MC carbides, Laves phases and δ needles were 1.3%, 0.6% and 4.4%, respectively. The secondary dendrite arm spacing was 157 μm ± 23 μm.

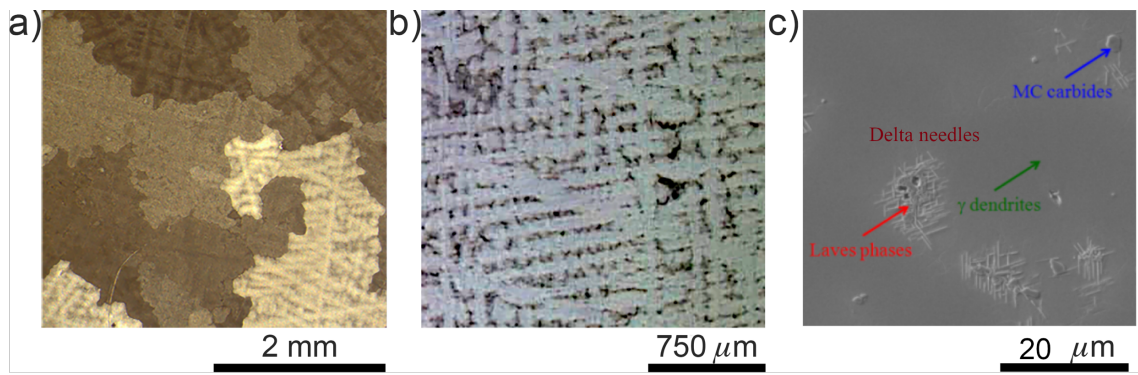

**Fig. 1. Microstructure of the IN718 alloy.** Optical micrographs of the as-cast IN718 alloy showing the (a) grain structure and (b) a detail of the dendrites inside a grain. (c) SEM micrograph showing common second-phase particles found in the interdendritic areas (MC carbides, δ ($Ni_3Nb$) needles and Laves phases)

The measurement area in each temper condition, of size 640 μm x 640 μm comprised one or a few grains, as shown in the EBSD maps of Fig 2.a, b and c for the solubilised, peak-aged and overaged precipitation states, respectively, where the colour code indicates the crystallographic direction of the surface normal in each case.

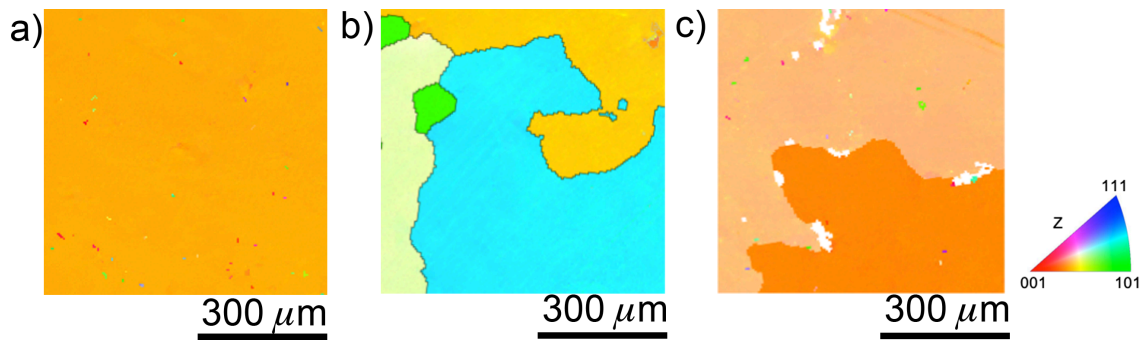

**Fig. 2. Crystallographic orientation maps.** EBSD maps showing crystallographic orientations of the grains present in the selected areas of the (a) solubilized, (b) peak-aged and (c) overaged precipitation states, respectively.

Due to the different diffusion rates of each chemical element, chemical segregations are expected in the cast material at the dendrite scale. Fig. 3 shows the compositional maps in the three representative areas for the seven most important alloying elements of this alloy: Ni, Cr, Fe, Mo, Al, Ti and Nb. The maps show that Ni presented a slight segregation towards the interdendritic spaces and that it was absent from the second-phase particles found in the interdendritic areas. On the other hand, since the diffusion of Cr and Fe is slower than other elements [7,8], they tended to be localized in the dendrites core region. In the case of Mo and Al, their content was lower than the other elements and were homogeneously distributed along the different microstructure features. On the contrary, Ti segregated strongly to the interdendritic areas, taking part on the formation of the second phases particles present in these regions, mainly MC carbides. Finally, Nb segregated towards the dendrites outer region, into the interdendritic area and the second-phase particles. As a result, and even though the average Nb content was 5.1 wt.%, the Nb content of the dendrite core was as low as 2 wt.%, while the Nb levels in the interdendritic regions, away from the second-phase particles, reached 8 wt.%. The Laves phases and δ needles appeared as islands containing about 25 wt.%Nb. No significant differences in terms of chemical segregation were found with temper condition, which indicates that the temperatures and times were not high or long enough, respectively, to homogenise the chemical composition.

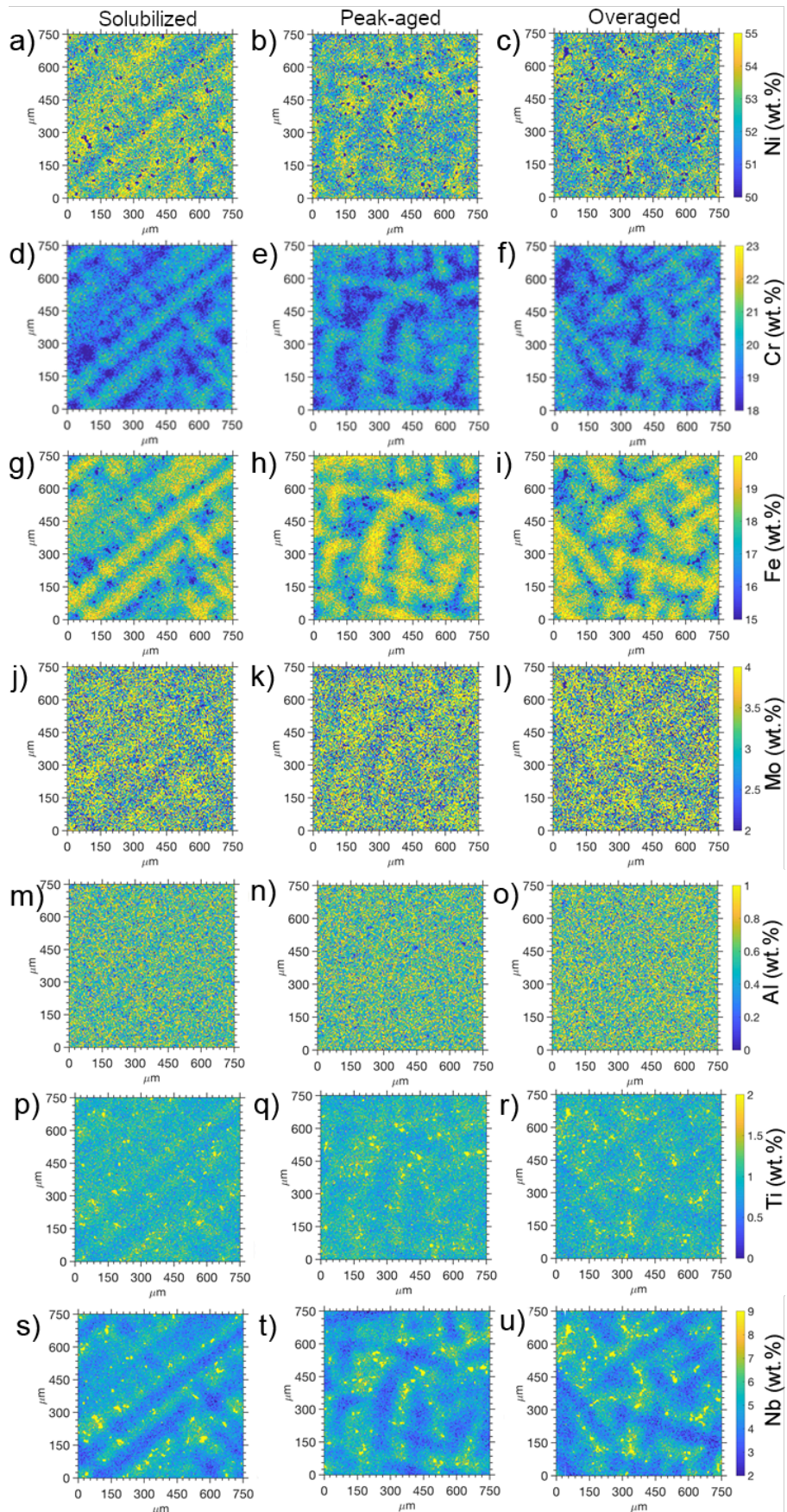

**Fig. 3. EDS compositional maps.** Local composition distribution values of Ni (a, b, c), Cr (d, e, f), Fe (g, h, i), Mo (j, k, l), Al (m, n, o), Ti (p, q, r) and Nb (s, t, u).

Finally, the precipitates in the peak aged condition were characterized by TEM. The alloy matrix comprises a FCC $\gamma$ phase consisting on a Ni based solid solution. The $\gamma$ matrix is strengthened by an intermetallic FCC $\gamma$' phase with composition $Ni_3(Ti,Al)$ and a BCT $\gamma$'' phase with composition $Ni_3Nb$ formed during ageing [9]. Fig.4 shows TEM images of the specimen in the peak aged condition at different magnifications. The $\gamma$'' precipitates exhibit an elongated disc shape, whereas the $\gamma$' precipitates are almost spheroidal. Both had an average size of 20 nm. The presence of complex precipitates comprised of two half-spheroidal $\gamma$' particles sandwiching a $\gamma$'' disc-shape precipitate could also be observed, as shown in Fig. 4b.

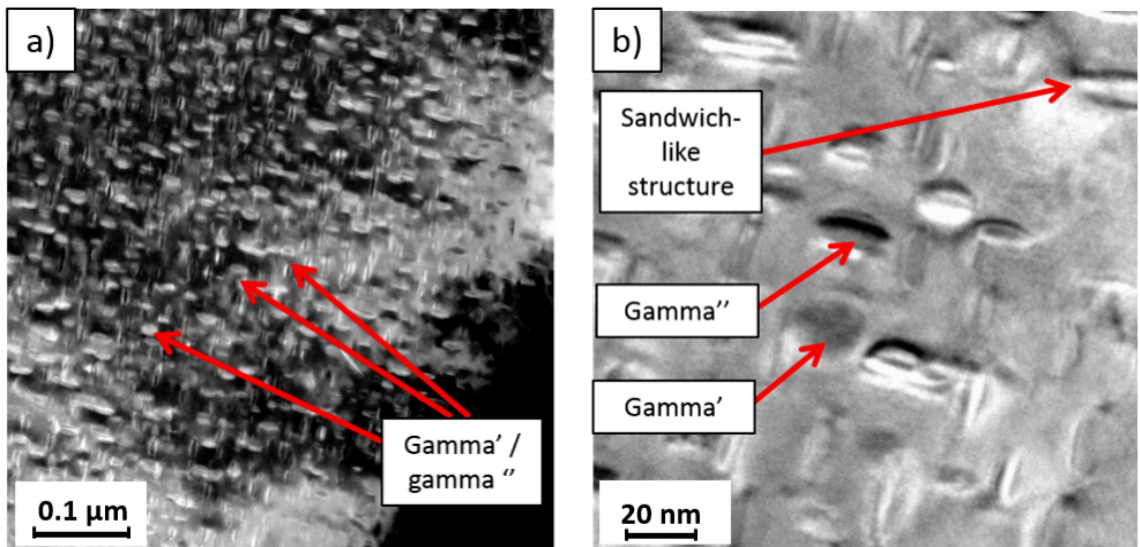

**Fig. 4.** TEM images showing the hardening γ' and γ'' precipitates in the peak aging condition

**2.2.- Mechanical properties**

The hardness (H) and reduced elastic modulus ($E_r$) maps for the solubilised, peak-aged and overaged precipitation state are shown in Fig. 5 a-c and d-f, respectively. The average hardness values, without considering the second phase particles, were 4.7, 7.8 and 6.2 GPa for the solubilised, peak-aged and overaged temper, respectively. However, the hardness values showed a strong gradient at the dendrite scale, with lower values within the dendrite cores that tend to increase towards the interdendritic regions (Fig. 5.a-c). The hardness difference between the dendrite core and the interdendritic region was as high as 2 GPa, depending on temper condition. Additionally, the second phases located in the interdendritic spaces, mainly MC carbides, presented the highest hardness values (up to 30 GPa).

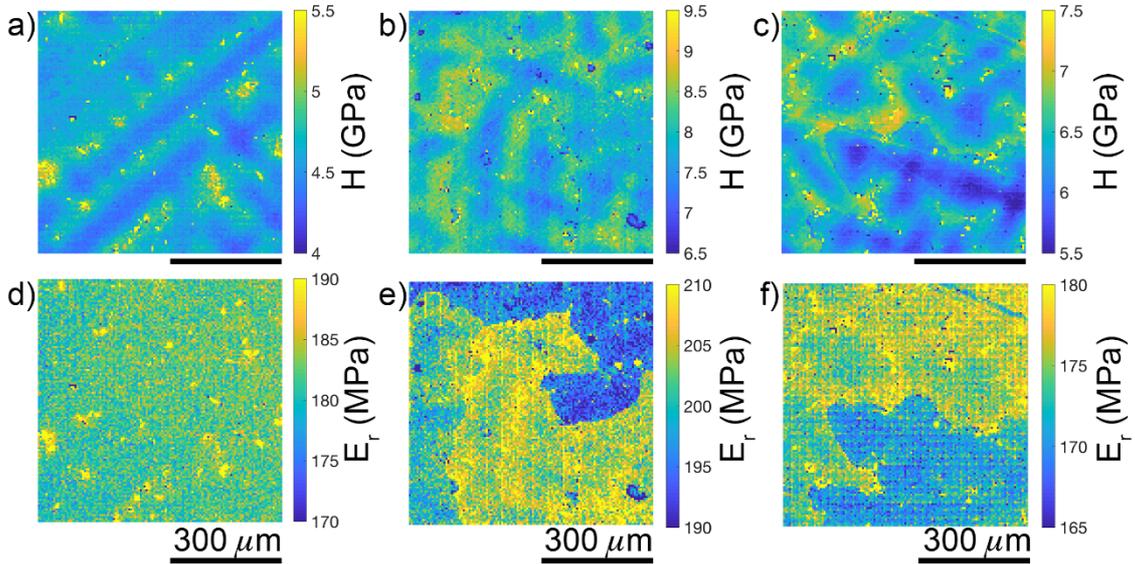

**Fig. 5. Mechanical properties maps.** Hardness (H) and reduced elastic modulus ($E_r$) maps obtained by XPM in the studied areas of the (a) solubilized, (b) peak-aged and (c) overaged precipitation states, respectively.

Regarding the reduced elastic modulus, Fig. 5.d, e and f, the values did not follow the trend observed in the hardness maps (Fig. 5.a-c) and they appear to be constant for each grain and less sensitive to chemical segregations at the dendrite scale. In the case of the solubilized state, the entire measurement area comprised just one grain with its surface normal oriented between the [001] and [101] directions (Fig. 2a), and the elastic modulus value was constant, around 182 MPa. Only the second phases present in the interdendritic spaces present much higher values, well over 200 MPa. On the other hand, the peak-aged and the overaged states present areas with different values of elastic modulus, that match closely with the different crystallographic orientations of the grains present in the measurement area in each case. The light blue grain in the peak-aged condition (Fig. 2b), with the surface normal between [101] and [111], presented the highest elastic modulus, around 210 MPa, (Fig. 5e), while the dark orange grain in the overaged condition (Fig. 2c), with its surface normal close to [001], presented the lowest elastic modulus, of around 165 MPa.

## 3.- Discussion

The correlative studied carried out showed that, even though cast IN718 typically shows grain sizes in the order of millimetres, is a very heterogeneous material at the microscale. The dendritic growth during solidification leads to a large heterogeneity at the scale of the SDAS (around 150 µm in this case). As a result, strong chemical segregations occur within each grain at the dendrite scale, with dendrite cores that are richer in Fe and Cr and depleted in Nb, while the interdendritic regions are much richer in Nb and in hard second phases, such as the TiC particles (Fig. 1). The chemical segregation through the dendrite radius occurs due to differences in the diffusion rates of each constitutive element. While the slower elements, Cr and Fe, remain in the dendrite cores (Fig. 3d-f and g-l, respectively), Ti and Nb tend to segregate towards the outer part of the dendrites (Fig. 3p-r and s-u, respectively), and specially into the interdendritic areas. In the particular case of Nb, the presence of this element is crucial for the formation of the metastable $\gamma''$-$Ni_3Nb$ precipitates responsible for the strengthening of Inconel 718. These precipitates are disc-shaped (Fig. 4) and play a key role as strengthening agents due to their higher volume fraction with respect to $\gamma'$ precipitates, as well as their coherency with the Ni matrix and the lattice distortion caused by the c-axis of the $D0_{22}$ body centred tetragonal $\gamma''$ structure [10,11]. Therefore, the Nb content is expected to play a crucial role on the local mechanical properties of this alloy. Fig. 6a represents the histogram of the Nb distribution for the three thermal treatments obtained from the corresponding EDS maps, excluding the second-phase particles (Fig. 3s-u). The distribution was similar for the three specimens, with a minimum and a maximum content of 2 and 8 wt.%, respectively, which indicates that Nb diffusion does not take place significantly during the temper treatments. On the contrary, the heat treatments are expected to affect the dissolution, precipitation and coarsening of the $\gamma''$ precipitates (Fig. 4), that are expected to be heterogeneously distributed at the dendrite scale, as a function of the local Nb content.

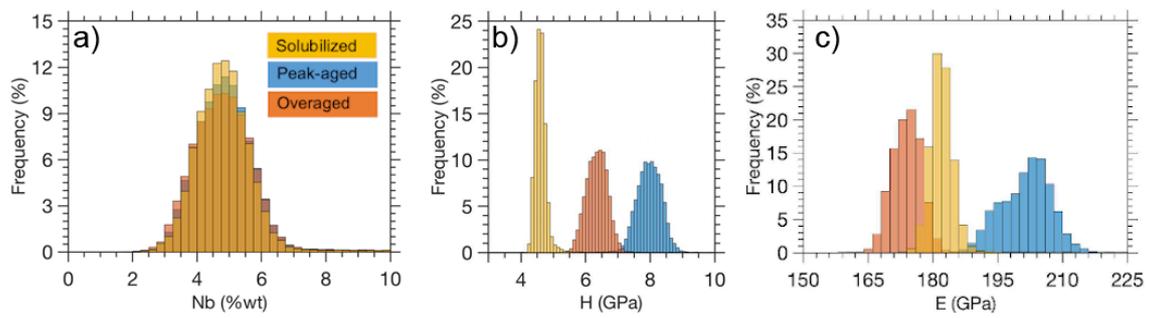

**Fig. 6. Nb concentration and mechanical properties values distributions.** (a) Values of the concentration of Nb, (b) hardness and (c) reduced elastic modulus in the three precipitation states.

In fact, the hardness maps (Fig. 5a, b and c) reveal a direct correlation with the Nb content (Fig. 3s, t and u) and the temper condition. The histogram of hardness values in Fig. 6b show an average hardness of 4.7, 7.8 and 6.2 GPa for the solubilised, peak-aged and overaged temper conditions, respectively, as expected from the precipitation state of the alloy. But more importantly, the hardness dispersion provides valuable information on the role of the Nb content in each case. For instance, the solubilised state presents a narrower hardness distribution than the peak-aged and the overaged states. In other words, the width of the hardness distribution in the solubilized state is about 1-1.2 GPa, while it increases to around 2 GPa in the peak-aged and the overaged states. This can be confirmed comparing the hardness values in the core and the surface of the dendrites arms in Fig. 5a,b and c. Nb is expected to be mainly in solid solution in the solubilised state, with a lower fraction of $\gamma''$ strengthening precipitates with respect to the peak aged and overaged conditions. This indicates that Nb segregation leads to larger mechanical property variations in $\gamma''$ strengthened than in solid solution strengthened Ni based superalloys. Moreover, the Nb content and the hardness maps, containing over 16000 data points in each case, represent an statistically significant dataset from which quantitative correlations can be extracted between hardness and Nb content for each temper condition, as shown in Fig. 7. This information represents a very valuable local tool to assess the quality of welds, repairs or 3D printed components, as the local hardness maps can be directly correlated to the local microstructure, to identify areas with a strong Nb segregation or with local tempers, as a result of heat affected zones.

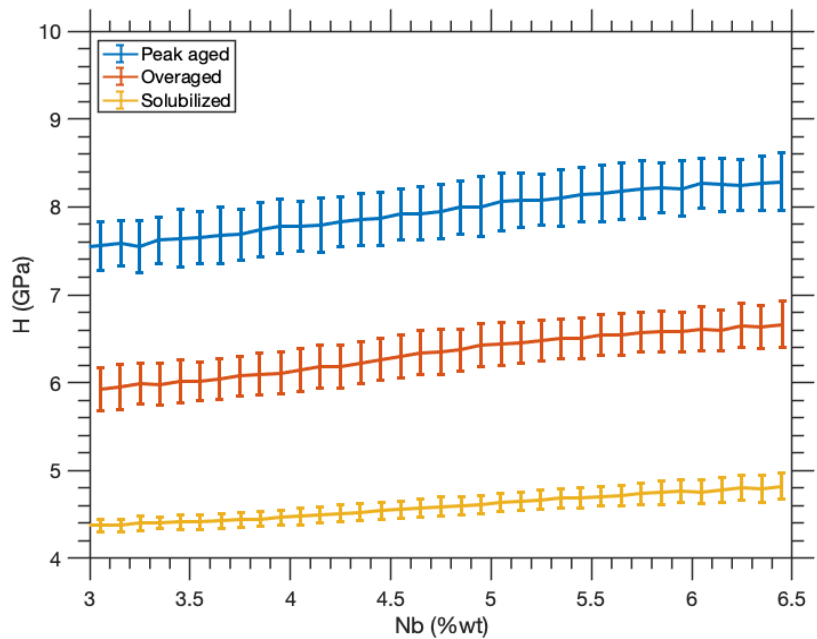

**Fig. 7.** Hardness versus Nb concentration for the solubilized, peak-aged and over-aged temper conditions.

Finally, it is interesting to notice that the reduced elastic modulus maps (Fig. 5d, e and f) do not show a correlation with the local chemical segregation or the temper condition. Instead, the elastic modulus maps present a direct correlation with the crystallographic orientation of the grains (Fig. 2a, b and c). This is not surprising considering the large elastic anisotropy of IN178 [12,13] and the fact that, contrary to hardness, elastic properties are relatively insensitive to microstructural features, such as precipitation stage or small variations in chemical composition. Fig. 8 plots the expected elastic modulus variation with crystallographic orientation [14], calculated using the single-crystal elastic constants of IN718 ($c_{11}$=259; $c_{12}$=179; $c_{44}$=109.6, in GPa) [12]. The elastic modulus is expected to be much higher in the <111> direction, 279 GPa, than in the <101> and <001> directions, 104 and 113 GPa, respectively. Even though nanoindentation imposes a complex stress state under the indent, it is interesting to notice that the relative values of indentation elastic modulus correlate well with the normal crystallographic orientation of the grains. For instance, the light blue grain in the peak aged condition in Fig. 2b, with the surface normal between [101] and [111], presents the highest elastic modulus, of around 210 GPa (Fig. 5e and 6c), while the dark orange grain in the overaged condition (Fig. 2c), with its surface normal close to [001], presents the lowest elastic modulus, of around 165 MPa (Fig. 5f and Fig.6c).

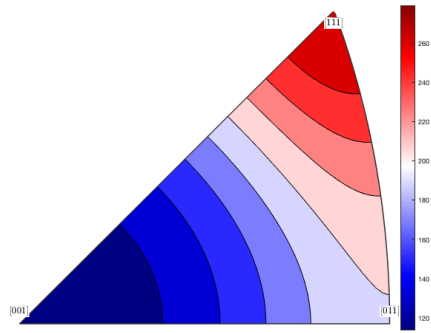

**Fig. 8.** Elastic modulus variation with crystallographic orientation expected for Inconel 718 [14] .

**4.- Conclusions**

A high-throughput correlative study of local mechanical properties, chemical composition and crystallographic orientation has been carried out in selected areas of cast Inconel 718 specimens subjected to three different tempers. The specimens showed a strong Nb segregation at the scale of the dendrite arms, with local Nb contents that varied between 2 wt.% in the core of the dendrite arms to 8 wt.% in the interdendritic regions and 25 wt.% within the second phase particles (MC carbides, Laves phases and δ phase needles). The nanohardness was found to correlate strongly with the local Nb content and the temper condition in each case. On the contrary, the indentation elastic moduli was not influenced by the local chemical composition or temper condition, but directly correlated with the grain orientation, due to the with high elastic anisotropy of nickel. Determining the correlation between the local mechanical properties, the chemical composition and the temper state can be a very valuable tool to assess the quality of cast components and other solidification processes, such as welds, repairs or 3D printed components. This way the local hardness maps can be directly correlated to the local

Paper submitted to Journal of Materials Research      12      2020

microstructure, to identify areas with a strong Nb segregation or with local tempers, as a result of heat affected zones.

**5.- Experimental methods**

5.1.- Sample preparation

Each sample was grinded using decreasing grit papers and then mechanically polished up to OP-S (0.025 µm). EBSD characterization required for an additional etching of the sample surface to resolve the Kikuchi´s patterns using Grundy´s reagent (52.6 cm$^3$ HCl, 36.9 cm$^3$ H$_2$O, 10.5 cm$^3$ HNO$_3$, 2.6 g CuCl$_2$, 2.6 g FeCl$_3$) during 60 s at room temperature.

5.2.- Microstructural characterization

Compositional characterization was carried out in the areas of interest by EDS using an Oxford detector mounted on a FEI FIB-FEGSEM Helios NanoLab 600i. We selected an acceleration voltage of 20 kV, a probe current of 2.7 nA and a working distance of 5 mm for mapping the concentration of Ni, Cr, Fe, Nb, Mo, Ti and Al. The maps were processed using the Oxford AZTec software selecting a 16 pixels binning window and presented as weight percentage (wt. %).

Crystallographic orientation mapping was performed in the same electron microscope using the Oxford Nordlys 2 EBSD detector working at 20 kV, a probe current of 2.7 nA, a working distance of 8 mm and a step size of 100 nm. The data was analysed using the *Oxford Instruments HKL Channel 5* software. Additionally, the precipitates were characterized by means of a transmission electron microscope (TEM) FEI Talos F200X, working at 200 kV.

5.3.- Mechanical characterization

The room temperature hardness and elastic modulus were determined using a Hysitron nanoindenter, model TI950, equipped with a 12 mN load cell and using a diamond Berkovich indenter with an apex angle of 142.35 ° and a tip radius of 150 nm. The XPM upgrade allows using the indenter piezoelectric actuator, instead of the stage movement, to move the indentation tip between indents, making the displacement process quicker and more accurate. The range of the piezoelectric actuator is 80 µm. In order to analyse the large areas representative of the cast microstructures, a 16 × 16 mosaic formed by rectangular grids of 8 × 8 indentations, was performed in each case. The indenter positioning within the grids was achieved by the piezoelectric actuator, while the movement between the rectangular grids to complete the mosaic was achieved through the conventional stage movement of the nanoindentation platform. Using this approach, each nanoindentation map was composed of a total of 16384 indents, and was obtained over a period of 12 hours. The indentation load was 10 mN in all cases, for which the contact area was in the range of 200 nm. The indents were separated by 5 µm to avoid any effect from the plastically deformed volume of previous indents [15]. This way, the total area analysed was 640 × 640 µm, with a resolution of 5 µm, covering several dendrites and second phases. Each indent consisted on a 0.1 s load-hold-unload cycle. The value of the hardness ($H$) was obtained from:

$$H = \frac{L_{max}}{A_{max}} \quad (1)$$

where $L_{max}$ is the maximum load reached during the test and $A_{max}$ is the contact area determined from the Oliver and Pharr method [6]. The reduced elastic modulus was obtained from the contact stiffness, i.e. the slope at the beginning of the unloading curve ($S=dP/dh$), using Eq. 2.

$$E_r = \frac{\sqrt{\pi}}{2\beta} \frac{S}{\sqrt{A_{max}}} \quad (2)$$

where $\beta$ is a dimensionless correction factor which accounts for the deviation in stiffness due to the lack of axisymmetry of the indenter tip with $\beta$ =1.02-1.19 for a triangular Berkovich punch [16]. The sample´s elastic modulus, $E_s$, can then be obtained using Eq. 3.

$$\frac{1}{E_r} = \frac{1-v_s^2}{E_s} + \frac{1-v_i^2}{E_i} \quad (3)$$

where E and $v$ are the elastic modulus and the Poisson´s ratio, respectively, of the sample (s) and of the indenter (i).

In order to ensure that the geometry of the indenter was not altered after such a large number of indents, the area function of the indenter was assessed before and after each indentation session. The diamond area function was determined by progressive loading-unloading cycles in a material with well-known mechanical properties, such as fused silica. No significant differences were found during the entire experimental campaign.

**Acknowledgements**

The financial support of MAT4.0-CM project funded by Madrid region under programme S2018/NMT-4381 is gratefully acknowledged. CG acknowledges funding from the Spanish Ministry of Science, Innovation and Universities (BES-2017-080201). We thank Marcos Jiménez for help with the TEM studies.

**Conflicts of interest**

The authors have no conflicts of interest to declare that are relevant to the content of this article.

**Data availability**

The datasets generated during and/or analysed during the current study are available from the corresponding author on reasonable request.